**Should the choice of BOIN design parameter *p.tox* only depend on the target DLT rate?**

Rong Lu, PhD*

*The Quantitative Sciences Unit, Division of Biomedical Informatics Research, Department of Medicine, Stanford University, Stanford, California

**Corresponding Author:** Rong Lu, PhD (https://orcid.org/0000-0003-4321-9144); ronglu@stanford.edu; 3180 Porter Drive, Palo Alto, CA 94304.

**Conflicts of Interest:** None

**Funding Sources:** This work is partially supported by the Biostatistics Shared Resource (BSR) of the NIH-funded Stanford Cancer Institute (P30CA124435) and the Stanford Center for Clinical and Translational Research and Education (UL1-TR003142).

**Role of the Funder/Sponsor:** The funders had no role in the design and conduct of the study; collection, management, analysis, and interpretation of the data; preparation, review, or approval of the manuscript; and decision to submit the manuscript for publication.

**KEY POINTS**

**Question:** When is setting the BOIN design parameter *p.tox* = 1.4 * *target.DLT.rate* not a great idea?

**Findings:** When the early stopping parameter *n.earlystop* is relatively small or the *cohortsize* value is not optimized via simulation, it might be better to use *p.tox* < 1.4 * *target.DLT.rate,* or try out different *cohortsize,* or increase *n.earlystop*, whichever is both feasible and provides better operating characteristics. This is because if the *cohortsize* was not optimized via simulation, even when *n.earlystop* = 12 and *cohortsize* > 3, the BOIN escalation/de-escalation rules generated using *p.tox* = 1.4 * *target.DLT.rate* could be exactly the same as those calculated using *p.tox* > 3 * *target.DLT.rate*, which might not be acceptable for some pediatric trials targeting 10% DLT rate..

**Meaning:** This study demonstrates the importance of interpreting BOIN design parameter *p.tox* as an interval of toxicity rates that are considered too toxic, rather than one prespecified value that corresponds to the lowest toxicity probability that is deemed overly toxic. When designing a dose-finding trial using BOIN, it is important to perform simulation studies to identify equivalent sets of BOIN design parameters that can generate the same boundary table so that we can better compare the safety properties of different boundary tables.

# ABSTRACT

**IMPORTANCE:** On December 10, 2021, the FDA published a Determination Letter, along with a Statistical Review and Evaluation Report, and concluded that under the non-informative prior, the local Bayesian optimal interval design (BOIN) design, in its revised form, can be designated fit-for-purpose for identifying the maximum tolerated dose (MTD) of a new drug, assuming that dose-toxicity relationship is monotonically increasing. Although setting the BOIN design parameter *p.tox* = 1.4 * *target.DLT.rate* is recommended in almost all BOIN methodology articles and is the default value in the R package *BOIN*, it's unclear if the choice of *p.tox* should only depend on the target DLT rate and whether certain range of p.tox could produce the same BOIN boundary table.

**DESIGN:** In this simulation study, following parameters were varied one at a time, using R package *BOIN*, to explore each parameter's effect on the equivalence intervals of *p.saf* and *p.tox*: 1) target DLT rate, 2) *n.earlystop*, 3) *cutoff.eli*, 4) *cohortsize*, and 5) *ncohort*. And a simple 3+3 design was used as an example to explore equivalent sets of BOIN design parameters that can generate the same boundary table.

**RESULTS:** When the early stopping parameter *n.earlystop* is relatively small or the *cohortsize* value is not optimized via simulation, it might be better to use p.tox < 1.4 * *target.DLT.rate*, or try out different cohort sizes, or increase *n.earlystop*, whichever is both feasible and provides better operating characteristics. This is because if the cohortsize was not optimized via simulation, even when *n.earlystop* = 12 and *cohortsize* > 3, the BOIN escalation/de-escalation rules generated using p.tox = 1.4 * *target.DLT.rate* could be exactly the same as those calculated using p.tox > 3 *

*target.DLT.rate*, which might not be acceptable for some pediatric trials targeting 10% DLT rate.

The traditional 3+3 design stops the dose finding process when 3 patients have been treated at the current dose level, 0 DLT has been observed, and the next higher dose has already been eliminated. If additional 3 patients were required to be treated at the current dose in the situation described above, the decision rules of this commonly used 3+3 design could be generated using BOIN design with target DLT rates ranging from 18% to 29%, *p.saf* ranging from 8% to 26%, and different *p.tox* values ranging from 39% to 99%. To generate this commonly used 3+3 design table, BOIN parameters also need to satisfy a set of conditions.

# INTRODUCTION

The Bayesian Optimal Interval (BOIN) design and its extensions are model-based designs that can guide both dose escalation and dose de-escalation in early phase trials [1-9]. The BOIN design was proposed to minimize the local decision error defined in section 2.2.1 of Liu and Yuan 2015 [1]. The R package *BOIN* published in August 2020 can be used to compare and implement BOIN designs for single-agent or drug-combination dose-finding trials [8]. When performing a BOIN design, it is expected that the total sample size budgeted will be much larger than that observed under the traditional 3+3 design without expansion cohorts. On December 10, 2021, the FDA published a Determination Letter [10], along with a Statistical Review and Evaluation Report [11], and concluded that under the non-informative prior, the local BOIN design, in its revised form, can be designated fit-for-purpose for identifying the maximum tolerated dose (MTD) of a new drug, assuming that the dose-toxicity relationship is monotonically increasing.

The BOIN framework allows the user to pre-specify target dose limiting toxicity (DLT) rate ($\phi$) as well as the following 8 design parameters [8]:

1. *ncohort*: The total number of cohorts.
2. *cohortsize*: The cohort size.
3. *n.earlystop*: The early stopping parameter. If the number of patients treated at the current dose reaches *n.earlystop*, stop the trial early and select the MTD based on the observed data. The default value of *n.earlystop* = 100 essentially turns off this type of early stopping.

4. *p.saf* ($\phi_1$): The highest toxicity probability that is deemed subtherapeutic (i.e., below the MTD) such that dose escalation should be made. The default value of *p.saf* = 0.6 * *target DLT rate*.

5. *p.tox* ($\phi_2$): The lowest toxicity probability that is deemed overly toxic such that dose de- escalation is required. The default value of *p.tox* = 1.4 * *target DLT rate*.

6. *cutoff.eli*: The cutoff to eliminate the overly toxic dose for safety. We recommend the default value *cutoff.eli* = 0.95 for general use.

7. *extrasafe*: Set *extrasafe* = TRUE to impose a stricter stopping rule.

8. *offset*: A small positive number (between 0 and 0.5) to control how strict the stopping rule is when *extrasafe* = TRUE. A larger value leads to a stricter stopping rule. The default value *offset* = 0.05 generally works well.

Under the non-informative prior, any BOIN design has an acceptable interval of observed toxicity rates ($\lambda_e$, $\lambda_d$] around its target DLT rate to determine whether the current dose is acceptable to retain [1, 8]:

$$\lambda_e = \frac{\log\left(\frac{1-\phi_1}{1-\phi}\right)}{\log\left\{\frac{\phi(1-\phi_1)}{\phi_1(1-\phi)}\right\}}, \qquad \lambda_d = \frac{\log\left(\frac{1-\phi}{1-\phi_2}\right)}{\log\left\{\frac{\phi_2(1-\phi)}{\phi(1-\phi_2)}\right\}}.$$

As with other model-based dose-finding algorithms, the operating characteristics of a BOIN design are greatly affected by the choice of its design parameters. Although setting the BOIN design parameter *p.tox* = 1.4 * *target.DLT.rate* is recommended in almost all BOIN methodology articles and is the default value in the R package *BOIN*, it's unclear why the choice of p.tox should only depend on the target DLT rate and whether certain range of p.tox could produce the same BOIN boundary table [1-9].

```
> get.boundary(target=0.1761482, ncohort=10, cohortsize=3, n.earlystop = 6,
+              p.saf = 0.1582749, p.tox = 0.892814, cutoff.eli = 0.8548338,
+              extrasafe = F)
$lambda_e
[1] 0.1670842

$lambda_d
[1] 0.556843

$boundary_tab

Number of patients treated 3 6
Escalate if # of DLT <=    0 1
Deescalate if # of DLT >=  2 2
Eliminate if # of DLT >=   2 2

$full_boundary_tab

Number of patients treated  1  2 3 4 5 6
Escalate if # of DLT <=     0  0 0 0 0 1
Deescalate if # of DLT >=   1  2 2 2 2 2
Eliminate if # of DLT >=    NA NA 2 2 2 2

attr(,"class")
[1] "boin"
Warning message:
In get.boundary(target = 0.1761482, ncohort = 10, cohortsize = 3,  :
  the value of n.earlystop is too low to ensure good operating characteristic
s. Recommend n.earlystop = 9 to 18.
```

**Figure 1a:** generating 3+3 design using BOIN design with target DLT rate = 17.6%

The traditional 3+3 design stops the dose finding process when 3 patients have been treated at the current dose level, 0 DLT has been observed, and the next higher dose has already been eliminated [12,13]. If additional 3 patients were required to be treated at the current dose in the situation described above, the 3+3 decision rules could be generated using different sets of BOIN parameters (Figure 1a & 1b):

```
> get.boundary(target=0.2903009, ncohort=10, cohortsize=3, n.earlystop = 6,
+              p.saf = 0.1501848, p.tox = 0.4812773, cutoff.eli = 0.6689076,
+              extrasafe = F)
$lambda_e
[1] 0.2146944

$lambda_d
[1] 0.3827505

$boundary_tab

Number of patients treated 3 6
Escalate if # of DLT <=    0 1
Deescalate if # of DLT >=  2 2
Eliminate if # of DLT >=   2 2

$full_boundary_tab

Number of patients treated  1  2 3 4 5 6
Escalate if # of DLT <=     0  0 0 0 1 1
Deescalate if # of DLT >=   1  1 2 2 2 2
Eliminate if # of DLT >=    NA NA 2 2 2 2

attr(,"class")
[1] "boin"
Warning message:
In get.boundary(target = 0.2903009, ncohort = 10, cohortsize = 3,  :
  the value of n.earlystop is too low to ensure good operating characteristics. Recommend n.earlystop = 9 to 18.
```

**Figure 1b:** generating 3+3 design using BOIN design with target DLT rate = 29%

In other words, the traditional 3+3 design is not a special case of BOIN design because it stops the dose finding process when 3 patients have been treated at the current dose level, 0 DLT has been observed, and the next higher dose has already been eliminated. However, if additional 3 patients were required to be treated at the current dose in the situation described above, this commonly used 3+3 design variant (traditional 3+3 design plus an expansion cohort of size 3) is a special case of BOIN design because

different sets of BOIN design parameters can generate its decision table below (Figure 1 & 2):

```
Number of patients treated    3 6
Escalate if # of DLT <=       0 1
Deescalate if # of DLT >=     2 2
Eliminate if # of DLT >=      2 2
```

**Table 1:** traditional 3+3 design with an expansion cohort of size 3

In this simulation study, following parameters will be varied one at a time, using R package *BOIN*, to explore each parameter's effect on the equivalence intervals of *p.saf* and *p.tox*: 1) target DLT rate, 2) *n.earlystop*, 3) *cutoff.eli*, 4) *cohortsize*, and 5) *ncohort*. And the 3+3 design boundary table above will be used as a simple example to explore all equivalent sets of BOIN design parameters that can generate the same boundary table.

**METHODS**

In this simulation study, all dose escalation/de-escalation boundary tables were calculated using the *get.bounddary()* function from R package *BOIN* [8]. *tryCatch()* function was used to handle *get.bounddary()* errors such as "the probability deemed safe cannot be higher than or too close to the target!" (line 13-23 in the example script 1a). All *p.saf* and *p.tox* values that produce the same BOIN boundary table (*i.e.* the same *$boundary_tab* output) are considered equivalent (line 72-163 in example script 1b). It's worth noting that while the *$boundary_tab* output is the same, the *$full_boundary_tab* outputs may be different (Figure 2b).

Equivalent *p.saf* and *p.tox* under varying target DLT rates:

Equivalent values of *p.saf* and *p.tox* were explored via uniform search under varying target DLT rates: *target* = 10%, 15%, 20%, 25%, 30%, 35%, or 40% (line 26 in the example script 1a) and fixed values of following design parameters (line 15-17 in the example script 1a):

- *ncohort* = 10
- *cohortsize* = 3
- *n.earlystop* = 12
- *cutoff.eli* = 95%
- *extrasafe* = FALSE

For each target DLT rate under evaluation, 100,000 pairs of *p.saf* and *p.tox* values were randomly drawn from the following uniform distributions (line 30-31, 55-56 in the example script 1a):

- *p.saf <- runif(1, min=0, max=target-0.0000001)*
- *p.tox <- runif(1, min=target+0.0000001, max=1)*

<u>Equivalent *p.saf* and *p.tox* under varying *n.earlystop*</u>:

Equivalent values of *p.saf* and *p.tox* were explored via uniform search under varying *n.earlystop* = 15, 18, 21, 24, 27, or 30 (line 26 in the example script 2a) and fixed values of following design parameters (line 15-17 in the example script 2a):

- *ncohort* = 10
- *cohortsize* = 3
- *target* = 10% (target DLT rate)
- *cutoff.eli* = 95%
- *extrasafe* = FALSE

For each *n.earlystop* value, 100,000 pairs of *p.saf* and *p.tox* values were randomly drawn from the following uniform distributions (line 30-31, 55-56 in the example script 2a):

- *p.saf <- runif(1, min=0, max=target-0.0000001)*
- *p.tox <- runif(1, min=target+0.0000001, max=1)*

<u>Equivalent *p.saf* and *p.tox* under varying *cutoff.eli*</u>:

Equivalent values of *p.saf* and *p.tox* were explored via uniform search under varying *cutoff.eli* = 70%, 80%, 90%, 97%, or 99% (line 26 in the example script 3a) and fixed values of following design parameters (line 15-17 in the example script 3a):

- *ncohort* = 10
- *cohortsize* = 3

- *target* = 10% (target DLT rate)
- *n.earlystop* = 12
- *extrasafe* = FALSE

For each *cutoff.eli* value, 100,000 pairs of *p.saf* and *p.tox* values were randomly drawn from the uniform distributions described in the previous sections (line 30-31, 55-56 in the [example script 3a](example script 3a)).

Equivalent *p.saf* and *p.tox* under varying *cohortsize*:

Equivalent values of *p.saf* and *p.tox* were explored via uniform search under varying *cohortsize* = 4, 5, 6, 7, or 8 (line 26 in the [example script 4a](example script 4a)) and fixed values of following design parameters (line 15-17 in the [example script 4a](example script 4a)):

- *ncohort* = 10
- *cutoff.eli* = 95%
- *target* = 10% (target DLT rate)
- *n.earlystop* = 12
- *extrasafe* = FALSE

For each *cohortsize*, 100,000 pairs of *p.saf* and *p.tox* values were randomly drawn from the uniform distributions described in the previous sections (line 30-31, 55-56 in the [example script 4a](example script 4a)).

Equivalent *p.saf* and *p.tox* under varying *ncohort*:

Equivalent values of *p.saf* and *p.tox* were explored via uniform search under varying *ncohort* = 5, 6, 7, 8, 9, 12, 20, 40, or 100 (line 23 in the example script 5a) and fixed values of following design parameters (line 12-14 in the example script 5a):

- *target* = 10% (target DLT rate)
- *cohortsize* = 3
- *n.earlystop* = 12
- *cutoff.eli* = 95%
- *extrasafe* = FALSE

For each *ncohort* value, 100,000 pairs of *p.saf* and *p.tox* values were randomly drawn from the uniform distributions described in the previous sections (line 27-28, 52-53 in the example script 5a).

Equivalent BOIN parameter sets for generating the same 3+3 design table

Equivalent values of *BOIN* design parameters were explored via uniform search under varying *offset* = 0.01, 0.05, 0.1, 0.2, 0.3, 0.4, or 0.49 (line 27 in the example script 6a) and fixed values of following design parameters (line 16-18 in the example script 6a):

- *cohortsize* = 3
- *ncohort* = 10
- *n.earlystop* = 6
- *extrasafe* = TRUE

For each *offset* value, 100,000 sets of *target* (target DLT rate), *p.saf*, *p.tox,* and *cutoff.eli* values were randomly drawn from the following uniform distributions (line 31-34, 57-60 in the example script 6a):

- *target <- runif(1,min=0,max=0.5)*
- *p.saf <- runif(1,min=0,max=target-0.0000001)*
- *p.tox <- runif(1,min=target+0.0000001,max=1)*
- *cutoff.eli <- runif(1,min=0,max=1)*

All sets of *BOIN* design parameters that can produce the 3+3 design table shown in Table 1 are considered equivalent (line 8, 65-68 in [example script 6a](#)):

To help clarify all methodological details and ensure reproducibility of this work, all simulation scripts and the outputs of these scripts are made available in a publicly accessible [code repository](#). These scripts can also be modified to facilitate simulation studies of other choices of BOIN design parameters. For example, simply replace the 3+3 design table in line 8 of [example script 6a](#) with any boundary table of interest, the updated [example script 6a](#) can be used to search all equivalent sets of *BOIN* design parameters that can produce this new boundary table of interest.

# RESULTS

## Results of equivalent *p.saf* and *p.tox* under different target DLT rates

When target DLT rate = 10%, *cohortsize* = 3, *ncohort* = 10, *n.earlystop* = 12, *cutoff.eli* = 95%, and *extrasafe* = FALSE, there are total 10 possible BOIN boundary tables, regardless of the choices of *p.saf* and *p.tox*. Equivalent intervals of *p.saf* and *p.tox* are visualized in **Figure 2a**, using different colors to indicate different boundary tables. In total, there are 2 equivalent intervals of *p.saf* in this setting: (0, 6.8%) and (6.8%, 9%), and 5 equivalent intervals of *p.tox*: (11%, 12.3%), (12.3%, 25.2%), (25.2%, 39%), (39%, 64.6%), and (64.6%, 99.9%). Therefore, when target DLT rate = 10%, *cohortsize* = 3, *ncohort* = 10, *n.earlystop* = 12, *cutoff.eli* = 95%, and *extrasafe* = FALSE, using *p.tox* = 1.4 \* *target.DLT.rate* to calculate BOIN boundary table is equivalent to using any p.tox $\in$ (12.3%, 25.2%), as long as *p.saf* values fall into one equivalent interval of *p.saf*. **Figure 2b** provides a few validation examples of the statement above. For detailed summary of the equivalent intervals reported above and their corresponding BOIN boundary tables, please see SupTable 2a.

When target DLT rate = 15%, *cohortsize* = 3, *ncohort* = 10, *n.earlystop* = 12, *cutoff.eli* = 95%, and *extrasafe* = FALSE, there are total 15 possible BOIN boundary tables, regardless of the choices of *p.saf* and *p.tox*. Equivalent intervals of *p.saf* and *p.tox* are visualized in SupFig 2c, using different colors to indicate different boundary tables. In total, there are 3 equivalent intervals of *p.saf* in this setting: (0, 3.9%), (3.9%, 7.9%) and (7.9%, 13.5%), and 5 equivalent intervals of *p.tox*: (16.5%, 18.4%), (18.4%, 30.8%), (30.8%, 37.2%), (37.2%, 56%), and (56%, 99.9%). Therefore, when target DLT rate =

15%, *cohortsize* = 3, *ncohort* = 10, *n.earlystop* = 12, *cutoff.eli* = 95%, and *extrasafe* = FALSE, using *p.tox = 1.4 * target.DLT.rate* to calculate BOIN boundary table is equivalent to using any p.tox ∈ (18.4%, 30.8%), as long as *p.saf* values fall into one equivalent interval of *p.saf*. And using *p.tox* = 33% to calculate BOIN boundary table is equivalent to using any p.tox ∈ (30.8%, 37.2%). For detailed summary of these equivalent intervals and their corresponding BOIN boundary tables, please see SupTable 2c.

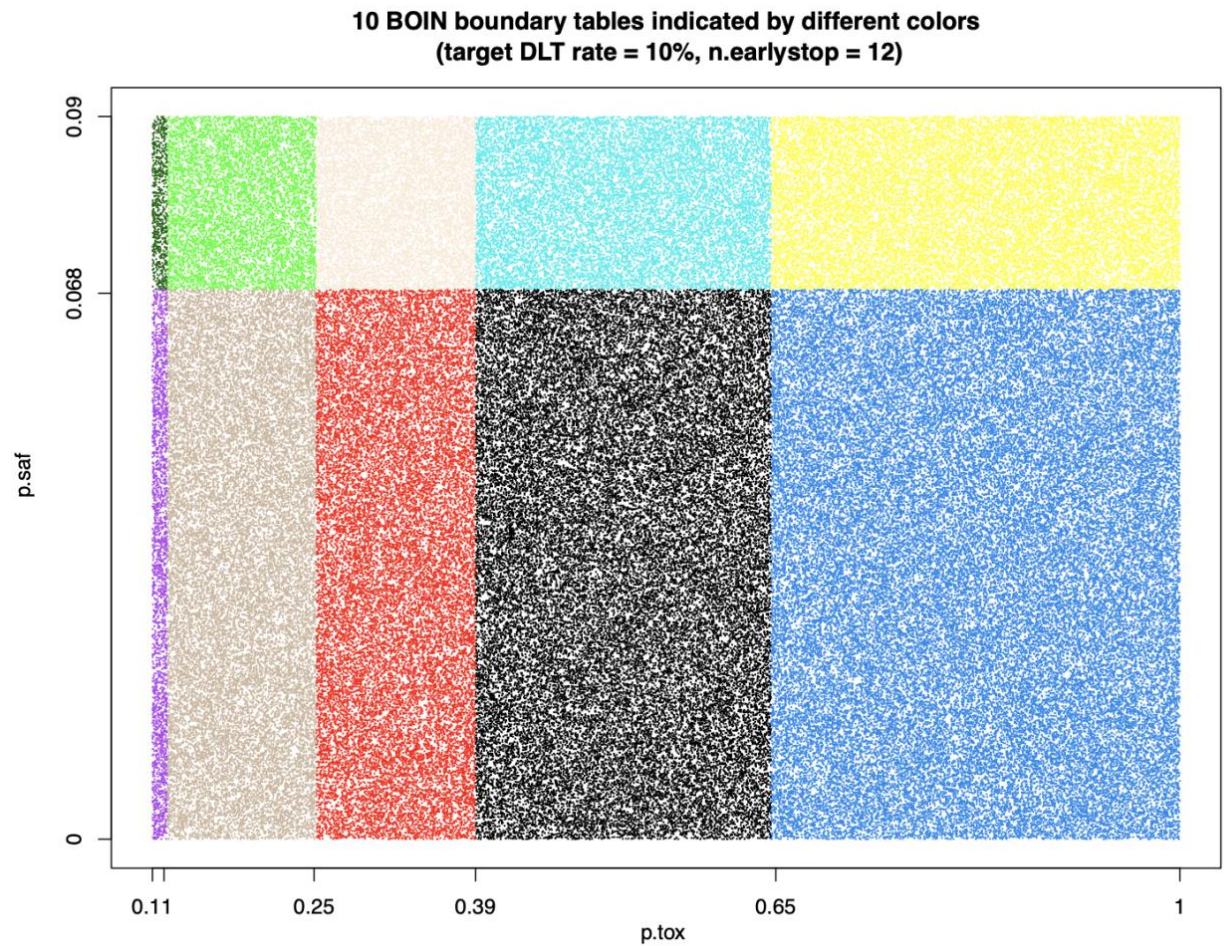

**Figure 2a:** equivalent intervals of *p.saf* and *p.tox* under target DLT rate = 10%, *cohortsize* = 3, *ncohort* = 10, *n.earlystop* = 12, *cutoff.eli* = 95%, and *extrasafe* = FALSE.

```
> get.boundary(target=0.1, ncohort=10, cohortsize=3, n.earlystop = 12,    > get.boundary(target=0.1, ncohort=10, cohortsize=3, n.earlystop = 12,    > get.boundary(target=0.1, ncohort=10, cohortsize=3, n.earlystop = 12,
+              p.saf = 0.00001, p.tox = 0.123, cutoff.eli = 0.95,        +              p.saf = 0.067, p.tox = 0.14, cutoff.eli = 0.95,          +              p.saf = 0.067, p.tox = 0.25, cutoff.eli = 0.95,
+              extrasafe = F)                                            +              extrasafe = F)                                          +              extrasafe = F)
$lambda_e                                                                $lambda_e                                                                $lambda_e
[1] 0.01130893                                                           [1] 0.08250041                                                           [1] 0.08250041

$lambda_d                                                                $lambda_d                                                                $lambda_d
[1] 0.1111531                                                            [1] 0.1190318                                                            [1] 0.1659562

$boundary_tab                                                            $boundary_tab                                                            $boundary_tab
Number of patients treated  3  6  9 12                                   Number of patients treated  3  6  9 12                                   Number of patients treated  3  6  9 12
Escalate if # of DLT <=     0  0  0  0                                   Escalate if # of DLT <=     0  0  0  0                                   Escalate if # of DLT <=     0  0  0  0
Deescalate if # of DLT >=   1  1  2  2                                   Deescalate if # of DLT >=   1  1  2  2                                   Deescalate if # of DLT >=   1  1  2  2
Eliminate if # of DLT >=    2  2  3  3                                   Eliminate if # of DLT >=    2  2  3  3                                   Eliminate if # of DLT >=    2  2  3  3

$full_boundary_tab                                                       $full_boundary_tab                                                       $full_boundary_tab
Number of patients treated  1  2  3  4  5  6  7  8  9 10 11 12           Number of patients treated  1  2  3  4  5  6  7  8  9 10 11 12           Number of patients treated  1  2  3  4  5  6  7  8  9 10 11 12
Escalate if # of DLT <=     0  0  0  0  0  0  0  0  0  0  0  0           Escalate if # of DLT <=     0  0  0  0  0  0  0  0  0  0  0  0           Escalate if # of DLT <=     0  0  0  0  0  0  0  0  0  0  0  0
Deescalate if # of DLT >=   1  1  1  1  1  1  1  1  2  2  2  2           Deescalate if # of DLT >=   1  1  1  1  1  1  1  1  2  2  2  2           Deescalate if # of DLT >=   1  1  1  1  1  1  2  2  2  2  2  2
Eliminate if # of DLT >=    NA NA 2  2  2  2  2  3  3  3  3  3           Eliminate if # of DLT >=    NA NA 2  2  2  2  2  3  3  3  3  3           Eliminate if # of DLT >=    NA NA 2  2  2  2  3  3  3  3  3  3

attr(,"class")                                                           attr(,"class")                                                           attr(,"class")
[1] "boin"                                                               [1] "boin"                                                               [1] "boin"
```

**Figure 2b:** When target DLT rate = 10%, *cohortsize* = 3, *ncohort* = 10, *n.earlystop* = 12, *cutoff.eli* = 95%, and *extrasafe* = FALSE, using *p.tox* = 1.4 * target.DLT.rate to calculate BOIN boundary table is equivalent to using any p.tox ∈ (12.3%, 25.2%), as long as *p.saf* values fall into one equivalent interval of *p.saf*: (0, 6.8%) or (6.8%, 9%).

When target DLT rate = 20%, *cohortsize* = 3, *ncohort* = 10, *n.earlystop* = 12, *cutoff.eli* = 95%, and *extrasafe* = FALSE, there are total 16 possible BOIN boundary tables, regardless of the choices of *p.saf* and *p.tox*. Equivalent intervals of *p.saf* and *p.tox* are visualized in [SupFig 2d](), using different colors to indicate different boundary tables. In total, there are 4 equivalent intervals of *p.saf* in this setting: (0, 2.2%), (2.2%, 5.1%), (5.1%, 13.7%) and (13.7%, 18%), and 4 equivalent intervals of *p.tox*: (22%, 24.6%), (24.6%, 30.5%), (30.5%, 48.8%), and (48.8%, 99.9%). Therefore, when target DLT rate = 20%, *cohortsize* = 3, *ncohort* = 10, *n.earlystop* = 12, *cutoff.eli* = 95%, and *extrasafe* = FALSE, using *p.tox* = 33% to calculate BOIN boundary table is equivalent to using any p.tox ∈ (30.5%, 48.8%), as long as *p.saf* values fall into one equivalent interval of *p.saf*. For detailed summary of these equivalent intervals and their corresponding BOIN boundary tables, please see [SupTable 2d]().

When target DLT rate = 25%, *cohortsize* = 3, *ncohort* = 10, *n.earlystop* = 12, *cutoff.eli* = 95%, and *extrasafe* = FALSE, there are total 30 possible BOIN boundary tables, regardless of the choices of *p.saf* and *p.tox*. Equivalent intervals of *p.saf* and *p.tox* are visualized in [SupFig 2e](), using different colors to indicate different boundary tables. In total, there are 5 equivalent intervals of *p.saf* in this setting: (0, 1.2%), (1.2%, 3.3%), (3.3%, 10.1%), (10.1%, 19.6%), and (19.6%, 22.5%), and 6 equivalent intervals of *p.tox*: (27.5%, 42.4%), (42.4%, 59.7%), (59.7%, 65.1%), (65.1%, 75%), (75%, 94.8%), and (94.8%, 99.9%). Therefore, when target DLT rate = 25%, *cohortsize* = 3, *ncohort* = 10, *n.earlystop* = 12, *cutoff.eli* = 95%, and *extrasafe* = FALSE, using *p.tox* = *1.4 \* target.DLT.rate* to calculate BOIN boundary table is equivalent to using any p.tox ∈

(27.5%, 42.4%), as long as *p.saf* values fall into one equivalent interval of *p.saf*. For detailed summary of these equivalent intervals and their corresponding BOIN boundary tables, please see SupTable 2e.

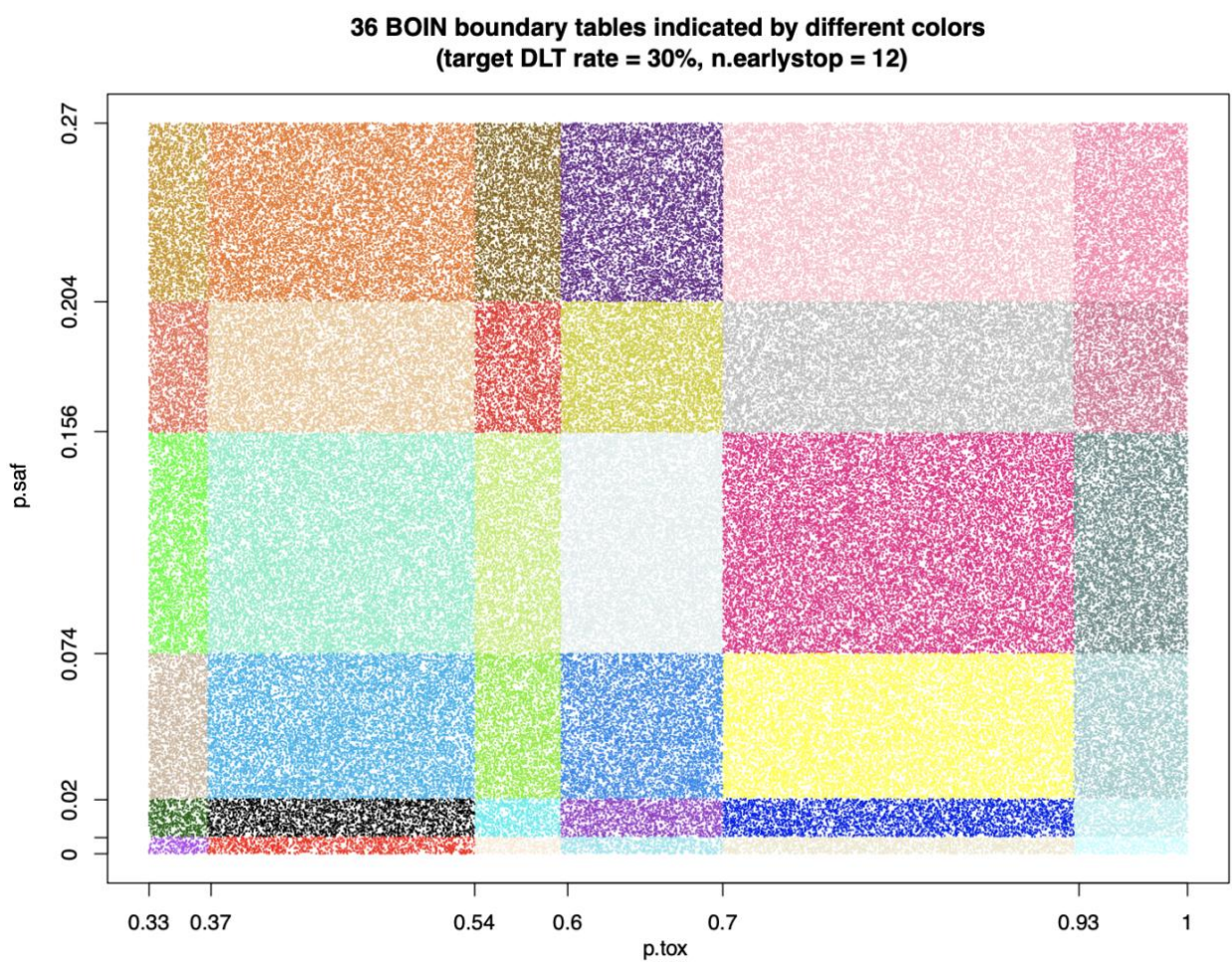

**Figure 2f:** equivalent intervals of *p.saf* and *p.tox* under target DLT rate = 30%, *cohortsize* = 3, *ncohort* = 10, *n.earlystop* = 12, *cutoff.eli* = 95%, and *extrasafe* = FALSE.

When target DLT rate = 30%, *cohortsize* = 3, *ncohort* = 10, *n.earlystop* = 12, *cutoff.eli* = 95%, and *extrasafe* = FALSE, there are total 36 possible BOIN boundary tables, regardless of the choices of *p.saf* and *p.tox*. Equivalent intervals of *p.saf* and *p.tox* are visualized in **Figure 2f**, using different colors to indicate different boundary tables. In total, there are 6 equivalent intervals of *p.saf* in this setting: (0, 0.6%), (0.6%, 2%), (2%,

7.4%), (7.4%, 15.6%), (15.6%, 20.4%), and (20.4%, 27%), and 6 equivalent intervals of *p.tox*: (33%, 36.8%), (36.8%, 54%), (54%, 59.6%), (59.6%, 70%), (70%, 92.7%), and (92.7%, 99.9%). Therefore, in this setting, using *p.tox = 1.4 * target.DLT.rate* to calculate BOIN boundary table is equivalent to using any p.tox ∈ (36.8%, 54%), as long as *p.saf* values fall into one equivalent interval of *p.saf*. For detailed summary of these equivalent intervals and their corresponding BOIN boundary tables, please see [SupTable 2f](). For results of equivalent intervals of *p.saf* and *p.tox* under target DLT rate = 35% and 40% in similar settings, please see [SupFig 2g](), [SupTable 2g](), [SupFig 2h](), and [SupTable 2h]().

Results of equivalent *p.saf* and *p.tox* under different *n.earlystop*

When target DLT rate = 10%, *cohortsize* = 3, *ncohort* = 10, *n.earlystop* = 15, *cutoff.eli* = 95%, and *extrasafe* = FALSE, there are total 21 possible BOIN boundary tables, regardless of the choices of *p.saf* and *p.tox*. Equivalent intervals of *p.saf* and *p.tox* are visualized in [SupFig 3a](), using different colors to indicate different boundary tables. In total, there are 3 equivalent intervals of *p.saf* in this setting: (0, 4.1%), (4.1%, 6.8%) and (6.8%, 9%), and 7 equivalent intervals of *p.tox*: (11%, 12.3%), (12.3%, 17.2%), (17.2%, 25.2%), (25.2%, 33.5%), (33.5%, 39%), (39%, 64.6%), and (64.6%, 99.9%). Therefore, in this setting, using *p.tox = 1.4 * target.DLT.rate* to calculate BOIN boundary table is equivalent to using any p.tox ∈ (12.3%, 17.2%), as long as *p.saf* values fall into one equivalent interval of *p.saf*. Compared to the results shown in **Figure 2a**, 4.1% is the only new interval boundary point added for *p.saf*; 17.2% and 33.5% are the two interval boundary points added for *p.tox*. The remaining boundary points of the equivalent intervals calculated under *n.earlystop* = 15 are the same as those calculated under *n.earlystop* = 12, for both *p.saf* and *p.tox*. SupTable 3a contains detailed summary of these equivalent intervals and their corresponding BOIN boundary tables.

When target DLT rate = 10%, *cohortsize* = 3, *ncohort* = 10, *n.earlystop* = 18, *cutoff.eli* = 95%, and *extrasafe* = FALSE, there are total 28 possible BOIN boundary tables, regardless of the choices of *p.saf* and *p.tox*. Equivalent intervals of *p.saf* and *p.tox* are visualized in SupFig 3b, using different colors to indicate different boundary tables. In total, there are 4 equivalent intervals of *p.saf* in this setting: (0, 2.6%), (2.6%, 4.1%), (4.1%, 6.8%) and (6.8%, 9%), and 7 equivalent intervals of *p.tox*. Compared to the results under *n.earlystop* = 15 above, 2.6% is the only new interval boundary point added for *p.saf*. The 7 equivalent intervals of *p.tox* calculated under *n.earlystop* = 18 are the same as those calculated under *n.earlystop* = 15. SupTable 3b contains detailed summary of these equivalent intervals and their corresponding BOIN boundary tables.

When target DLT rate = 10%, *cohortsize* = 3, *ncohort* = 10, *n.earlystop* = 21, *cutoff.eli* = 95%, and *extrasafe* = FALSE, there are total 45 possible BOIN boundary tables, regardless of the choices of *p.saf* and *p.tox*. Equivalent intervals of *p.saf* and *p.tox* are visualized in SupFig 3c, using different colors to indicate different boundary tables. In total, there are 5 equivalent intervals of *p.saf* in this setting: (0, 1.7%), (1.7%, 2.6%), (2.6%, 4.1%), (4.1%, 6.8%) and (6.8%, 9%), and 9 equivalent intervals of *p.tox*: (11%, 12.3%), (12.3%, 17.2%), (17.2%, 19.4%), (19.4%, 25.2%), (25.2%, 31.1%), (31.1%, 33.5%), (33.5%, 39%), (39%, 64.6%), and (64.6%, 99.9%). Compared to the results

under *n.earlystop* = 18 above, 1.7% is the only new interval boundary point added for *p.saf*; 19.4% and 31.1% are the two interval boundary points added for *p.tox*. The remaining boundary points of the equivalent intervals calculated under *n.earlystop* = 21 are the same as those calculated under *n.earlystop* = 18, for both *p.saf* and *p.tox*. SupTable 3c contains detailed summary of these equivalent intervals and their corresponding BOIN boundary tables. For results of equivalent intervals of *p.saf* and *p.tox* under *n.earlystop* = 24 in similar settings, please see SupFig 3d, and SupTable 3d.

Results of equivalent *p.saf* and *p.tox* under different *cutoff.eli*

When target DLT rate = 10%, *cohortsize* = 3, *ncohort* = 10, *n.earlystop* = 12, *cutoff.eli* = 97%, and *extrasafe* = FALSE, there are total 12 possible BOIN boundary tables, regardless of the choices of *p.saf* and *p.tox*. Equivalent intervals of *p.saf* and *p.tox* are visualized in SupFig 4a, using different colors to indicate different boundary tables. In total, there are 2 equivalent intervals of *p.saf* in this setting: (0, 6.8%) and (6.8%, 9%), and 6 equivalent intervals of *p.tox*: (11%, 12.3%), (12.3%, 25.2%), (25.2%, 39%), (39%, 45.8%), (45.8%, 64.6%), and (64.6%, 99.9%). Therefore, in this setting, using *p.tox* = 1.4 * *target.DLT.rate* to calculate BOIN boundary table is equivalent to using any p.tox ∈ (12.3%, 25.2%), as long as *p.saf* values fall into one equivalent interval of *p.saf*. Compared to the results shown in **Figure 2a**, no new interval boundary point is added for *p.saf*; 45.8% is the only interval boundary point added for *p.tox*. The remaining boundary points of the equivalent intervals calculated under *cutoff.eli* = 97% are the same as those calculated under *cutoff.eli* = 95% for both *p.saf* and *p.tox*. SupTable 4a contains detailed summary of these equivalent intervals and their corresponding BOIN boundary tables. When target DLT rate = 10%, *cohortsize* = 3, *ncohort* = 10, *n.earlystop* = 12, *cutoff.eli* = 99%, and *extrasafe* = FALSE, all equivalent interval results are the same as those calculated under *cutoff.eli* = 97% if other design parameters stay the same (SupFig 4a, SupTable 4a).

When target DLT rate = 10%, *cohortsize* = 3, *ncohort* = 10, *n.earlystop* = 12, *cutoff.eli* = 90%, and *extrasafe* = FALSE, there are total 6 possible BOIN boundary tables, regardless of the choices of *p.saf* and *p.tox*. Equivalent intervals of *p.saf* and *p.tox* are visualized in SupFig 4b, using different colors to indicate different boundary tables. In total, there are 2 equivalent intervals of *p.saf* in this setting: (0, 6.8%) and (6.8%, 9%), and 3 equivalent intervals of *p.tox*: (11%, 12.3%), (12.3%, 25.2%), and (25.2, 99.9%). So, in this setting, using *p.tox* = *1.4 \* target.DLT.rate* to calculate BOIN boundary table is equivalent to using any p.tox ∈ (12.3%, 25.2%), as long as *p.saf* values fall into one equivalent interval of *p.saf*. Compared to the results shown in **Figure 2a**, no new interval boundary point is added or reduced for *p.saf*; 39% and 64.6% are the 2 interval boundary points removed for *p.tox*. The remaining boundary points of the equivalent intervals calculated under *cutoff.eli* = 90% are the same as those calculated under *cutoff.eli* = 95% for both *p.saf* and *p.tox*. SupTable 4b contains detailed summary of these equivalent intervals and their corresponding BOIN boundary tables.

When target DLT rate = 10%, *cohortsize* = 3, *ncohort* = 10, *n.earlystop* = 12, *cutoff.eli* = 80%, and *extrasafe* = FALSE, there are total 4 possible BOIN boundary tables, regardless of the choices of *p.saf* and *p.tox*. Equivalent intervals of *p.saf* and *p.tox* are visualized in SupFig 4c, using different colors to indicate different boundary tables. In total, there are 2 equivalent intervals of *p.saf* in this setting: (0, 6.8%) and (6.8%, 9%), and 2 equivalent intervals of *p.tox*: (11%, 12.3%) and (12.3%, 99.9%). So, in this setting, using *p.tox = 1.4 * target.DLT.rate* to calculate BOIN boundary table is equivalent to using any p.tox ∈ (12.3%, 99.9%), as long as *p.saf* values fall into one equivalent interval of *p.saf*. This is an example of why *cutoff.eli* < 90% should be used with caution. SupTable 4c contains detailed summary of these equivalent intervals and their corresponding BOIN boundary tables.

When target DLT rate = 10%, *cohortsize* = 3, *ncohort* = 10, *n.earlystop* = 12, *cutoff.eli* = 70%, and *extrasafe* = FALSE, there are only 2 possible BOIN boundary tables, regardless of the choices of *p.saf* and *p.tox*. Equivalent intervals of *p.saf* and *p.tox* are visualized in SupFig 4d, using different colors to indicate different boundary tables. The 2 equivalent intervals of *p.saf* in this setting are (0, 6.8%) and (6.8%, 9%), and the only equivalent interval of *p.tox* is (11%, 99.9%). So, in this setting, using *p.tox = 1.4 * target.DLT.rate* to calculate BOIN boundary table is equivalent to using any valid values of p.tox, as long as *p.saf* values fall into one equivalent interval of *p.saf*. This is again an example of why *cutoff.eli* < 90% should be used with caution. SupTable 4d contains detailed summary of these equivalent intervals and their corresponding BOIN boundary tables.

Results of equivalent *p.saf* and *p.tox* under varying *cohortsize*

When target DLT rate = 10%, *cohortsize* = 4, *ncohort* = 10, *n.earlystop* = 12, *cutoff.eli* = 95%, and *extrasafe* = FALSE, there are total 8 possible BOIN boundary tables, regardless of the choices of *p.saf* and *p.tox*. Equivalent intervals of *p.saf* and *p.tox* are visualized in SupFig 5a, using different colors to indicate different boundary tables. In total, there are 2 equivalent intervals of *p.saf* in this setting: (0, 6.8%) and (6.8%, 9%), and 4 equivalent intervals of *p.tox*: (11%, 15.3%), (15.3%, 25.2%), (25.2%, 45.8%), and (45.8%, 99.9%). Therefore, in this setting, using *p.tox = 33%* to calculate BOIN boundary table is equivalent to using any p.tox ∈ (25.2%, 45.8%), as long as *p.saf* values fall into one equivalent interval of *p.saf*. Compared to the results shown in **Figure 2a**, equivalent intervals of *p.saf* are the same, but number of equivalent intervals of *p.tox* reduced from 5 to 4. SupTable 5a contains detailed summary of these equivalent intervals and their corresponding BOIN boundary tables.

When target DLT rate = 10%, *cohortsize* = 5, *ncohort* = 10, *n.earlystop* = 12, *cutoff.eli* = 95%, and *extrasafe* = FALSE, there are total 2 possible BOIN boundary tables, regardless of the choices of *p.saf* and *p.tox*. Equivalent intervals of *p.saf* and *p.tox* are visualized in SupFig 5b, using different colors to indicate different boundary tables. In total, there is only one equivalent interval of *p.saf* (0, 9%), and 2 equivalent intervals of *p.tox*: (11%, 33.5%) and (33.5%, 99.9%). Therefore, in this setting, using *p.tox = 1.4 \* target.DLT.rate* to calculate BOIN boundary table is equivalent to using *p.tox > 3 \* target.DLT.rate* in this setting. **Figure 5b** provides 3 validation examples of the statement above. SupTable 5b contains detailed summary of these equivalent intervals and their corresponding BOIN boundary tables.

```
> get.boundary(target=0.1, ncohort=10, cohortsize=5, n.earlystop = 12,
+              p.saf = 0.09, p.tox = 0.115, cutoff.eli = 0.95,
+              extrasafe = F)
$lambda_e
[1] 0.09492142

$lambda_d
[1] 0.1073464

$boundary_tab
Number of patients treated  5 10
Escalate if # of DLT <=     0  0
Deescalate if # of DLT >=   1  2
Eliminate if # of DLT >=    2  3

$full_boundary_tab
Number of patients treated   1  2 3 4 5 6 7 8 9 10 11 12
Escalate if # of DLT <=      0  0 0 0 0 0 0 0 0  0  1  1
Deescalate if # of DLT >=    1  1 1 1 1 1 1 1 1  2  2  2
Eliminate if # of DLT >=    NA NA 2 2 2 2 3 3 3  3  3  3

attr(,"class")
[1] "boin"
```

```
> get.boundary(target=0.1, ncohort=10, cohortsize=5, n.earlystop = 12,
+              p.saf = 0.067, p.tox = 0.14, cutoff.eli = 0.95,
+              extrasafe = F)
$lambda_e
[1] 0.08250041

$lambda_d
[1] 0.1190318

$boundary_tab
Number of patients treated  5 10
Escalate if # of DLT <=     0  0
Deescalate if # of DLT >=   1  2
Eliminate if # of DLT >=    2  3

$full_boundary_tab
Number of patients treated   1  2 3 4 5 6 7 8 9 10 11 12
Escalate if # of DLT <=      0  0 0 0 0 0 0 0 0  0  0  0
Deescalate if # of DLT >=    1  1 1 1 1 1 1 1 2  2  2  2
Eliminate if # of DLT >=    NA NA 2 2 2 2 3 3 3  3  3  3

attr(,"class")
[1] "boin"
```

```
> get.boundary(target=0.1, ncohort=10, cohortsize=5, n.earlystop = 12,
+              p.saf = 0.00001, p.tox = 0.33, cutoff.eli = 0.95,
+              extrasafe = F)
$lambda_e
[1] 0.01130893

$lambda_d
[1] 0.1981929

$boundary_tab
Number of patients treated  5 10
Escalate if # of DLT <=     0  0
Deescalate if # of DLT >=   1  2
Eliminate if # of DLT >=    2  3

$full_boundary_tab
Number of patients treated   1  2 3 4 5 6 7 8 9 10 11 12
Escalate if # of DLT <=      0  0 0 0 0 0 0 0 0  0  0  0
Deescalate if # of DLT >=    1  1 1 1 2 2 2 2 2  3  3  3
Eliminate if # of DLT >=    NA NA 2 2 2 2 3 3 3  3  3  3

attr(,"class")
[1] "boin"
```

**Figure 5b:** When target DLT rate = 10%, *cohortsize* = 5, *ncohort* = 10, *n.earlystop* = 12, *cutoff.eli* = 95%, and *extrasafe* = FALSE, using *p.tox* = 1.4 * target.DLT.rate to calculate BOIN boundary table is equivalent to using *p.tox* > 3 * target.DLT.rate.

When target DLT rate = 10%, *cohortsize* = 6, *ncohort* = 10, *n.earlystop* = 12, *cutoff.eli* = 95%, and *extrasafe* = FALSE, there are total 4 possible BOIN boundary tables, regardless of the choices of *p.saf* and *p.tox*. Equivalent intervals of *p.saf* and *p.tox* are visualized in SupFig 5c, using different colors to indicate different boundary tables. In total, there are 2 equivalent intervals of *p.saf* in this setting: (0, 6.8%) and (6.8%, 9%), and 2 equivalent intervals of *p.tox*: (11%, 25.2%) and (25.2%, 99.9%). Therefore, in this setting, using *p.tox* = *33%* to calculate BOIN boundary table is equivalent to using *p.tox* = *99%*, as long as *p.saf* values fall into one equivalent interval of *p.saf*. **Figure 5c** provides 3 validation examples of the statement above. Compared to the results shown in **Figure 2a**, equivalent intervals of *p.saf* are the same, but number of equivalent intervals of *p.tox* reduced from 5 to 2. SupTable 5c contains detailed summary of these equivalent intervals and their corresponding BOIN boundary tables.

Results of equivalent *p.saf* and *p.tox* under different *ncohort*

Although *ncohort* is one of the required input parameter of *get.boundary()*, the choice of *ncohort* value has no effect on the calculation of BOIN boundary table. For example, when target DLT rate = 10%, *cohortsize* = 3, *n.earlystop* = 12, *cutoff.eli* = 95%, and *extrasafe* = FALSE, there are the same 10 possible BOIN boundary tables, regardless of the choices of *p.saf*, *p.tox,* and *ncohort* ∈ {5, 6, 7, 8, 9, 10, 12, 20, 40, 100}. **Figure 6c** also provides 3 validation examples. The main purpose of specifying *ncohort* is to terminate dose finding process when the sample size budget is reached.

Results of equivalent BOIN parameter sets for generating the same 3+3 rules

```
> get.boundary(target=0.1, ncohort=10, cohortsize=6, n.earlystop = 12,    > get.boundary(target=0.1, ncohort=10, cohortsize=6, n.earlystop = 12,    > get.boundary(target=0.1, ncohort=10, cohortsize=6, n.earlystop = 12,
+              p.saf = 0.068, p.tox = 0.252, cutoff.eli = 0.95,            +              p.saf = 0.00001, p.tox = 0.33, cutoff.eli = 0.95,           +              p.saf = 0.00001, p.tox = 0.99, cutoff.eli = 0.95,
+              extrasafe = F)                                              +              extrasafe = F)                                              +              extrasafe = F)
$lambda_e                                                                  $lambda_e                                                                  $lambda_e
[1] 0.08306706                                                             [1] 0.01130893                                                             [1] 0.01130893

$lambda_d                                                                  $lambda_d                                                                  $lambda_d
[1] 0.1667718                                                              [1] 0.1981929                                                              [1] 0.6624826

$boundary_tab                                                              $boundary_tab                                                              $boundary_tab

Number of patients treated 6 12                                            Number of patients treated 6 12                                            Number of patients treated 6 12
Escalate if # of DLT <=    0  0                                            Escalate if # of DLT <=    0  0                                            Escalate if # of DLT <=    0  0
Deescalate if # of DLT >=  2  3                                            Deescalate if # of DLT >=  2  3                                            Deescalate if # of DLT >=  2  3
Eliminate if # of DLT >=   2  3                                            Eliminate if # of DLT >=   2  3                                            Eliminate if # of DLT >=   2  3

$full_boundary_tab                                                         $full_boundary_tab                                                         $full_boundary_tab

Number of patients treated  1  2 3 4 5 6 7 8 9 10 11 12                    Number of patients treated  1  2 3 4 5 6 7 8 9 10 11 12                    Number of patients treated  1  2 3 4 5 6 7 8 9 10 11 12
Escalate if # of DLT <=     0  0 0 0 0 0 0 0 0  0  0  0                    Escalate if # of DLT <=     0  0 0 0 0 0 0 0 0  0  0  0                    Escalate if # of DLT <=     0  0 0 0 0 0 0 0 0  0  0  0
Deescalate if # of DLT >=   1  1 1 1 1 2 2 2 2  2  2  3                    Deescalate if # of DLT >=   1  1 1 1 1 2 2 2 2  2  3  3                    Deescalate if # of DLT >=   1  2 2 2 2 2 3 3  3  3  3
Eliminate if # of DLT >=   NA NA 2 2 2 2 3 3 3  3  3  3                    Eliminate if # of DLT >=   NA NA 2 2 2 2 3 3 3  3  3  3                    Eliminate if # of DLT >=   NA NA 2 2 2 2 3 3  3  3  3

attr(,"class")                                                             attr(,"class")                                                             attr(,"class")
[1] "boin"                                                                 [1] "boin"                                                                 [1] "boin"
```

**Figure 5c:** When target DLT rate = 10%, *cohortsize* = 6, *ncohort* = 10, *n.earlystop* = 12, *cutoff.eli* = 95%, and *extrasafe* = FALSE, using *p.tox* = 33% to calculate BOIN boundary table is equivalent to using *p.tox* = 99%, as long as *p.saf* values fall into one equivalent interval of *p.saf*.

```
> get.boundary(target=0.1, ncohort=5, cohortsize=3, n.earlystop = 12,    > get.boundary(target=0.1, ncohort=10, cohortsize=3, n.earlystop = 12,   > get.boundary(target=0.1, ncohort=10000, cohortsize=3, n.earlystop = 12,
+           p.saf = 0.068, p.tox = 0.14, cutoff.eli = 0.95,              +           p.saf = 0.00001, p.tox = 0.14, cutoff.eli = 0.95,             +           p.saf = 0.00001, p.tox = 0.14, cutoff.eli = 0.95,
+           extrasafe = F)                                               +           extrasafe = F)                                                +           extrasafe = F)
$lambda_e                                                                 $lambda_e                                                                  $lambda_e
[1] 0.08306706                                                            [1] 0.01130893                                                             [1] 0.01130893

$lambda_d                                                                 $lambda_d                                                                  $lambda_d
[1] 0.1190318                                                             [1] 0.1190318                                                              [1] 0.1190318

$boundary_tab                                                             $boundary_tab                                                              $boundary_tab

Number of patients treated  3 6 9 12                                      Number of patients treated  3 6 9 12                                       Number of patients treated  3 6 9 12
Escalate if # of DLT <=     0 0 0  0                                      Escalate if # of DLT <=     0 0 0  0                                       Escalate if # of DLT <=     0 0 0  0
Deescalate if # of DLT >=   1 1 2  2                                      Deescalate if # of DLT >=   1 1 2  2                                       Deescalate if # of DLT >=   1 1 2  2
Eliminate if # of DLT >=    2 2 3  3                                      Eliminate if # of DLT >=    2 2 3  3                                       Eliminate if # of DLT >=    2 2 3  3

$full_boundary_tab                                                        $full_boundary_tab                                                         $full_boundary_tab

Number of patients treated  1  2 3 4 5 6 7 8 9 10 11 12                   Number of patients treated  1  2 3 4 5 6 7 8 9 10 11 12                    Number of patients treated  1  2 3 4 5 6 7 8 9 10 11 12
Escalate if # of DLT <=     0  0 0 0 0 0 0 0 0  0  0  0                   Escalate if # of DLT <=     0  0 0 0 0 0 0 0 0  0  0  0                    Escalate if # of DLT <=     0  0 0 0 0 0 0 0 0  0  0  0
Deescalate if # of DLT >=   1  1 1 1 1 1 1 1 2  2  2  2                   Deescalate if # of DLT >=   1  1 1 1 1 1 1 1 2  2  2  2                    Deescalate if # of DLT >=   1  1 1 1 1 1 1 1 2  2  2  2
Eliminate if # of DLT >=    NA NA 2 2 2 2 3 3 3  3  3  3                  Eliminate if # of DLT >=    NA NA 2 2 2 2 3 3 3  3  3  3                   Eliminate if # of DLT >=    NA NA 2 2 2 2 3 3 3  3  3  3

attr(,"class")                                                            attr(,"class")                                                             attr(,"class")
[1] "boin"                                                                [1] "boin"                                                                 [1] "boin"
```

**Figure 6c:** Although *ncohort* is one of the required input parameter of *get.boundary()*, the choice of *ncohort* value has no effect on the calculation of BOIN boundary table.

The random search script described in the method section sampled 8,127 sets of BOIN parameters that can generate the 3+3 design table shown in **Table 1**. These 8,127 sets of BOIN parameter values are all listed in SupTable 7. These 8,127 sets of BOIN parameter values satisfy following conditions:

- $0.17 < target.DLT.rate < 0.30$
- $0.08 < p.saf < 0.26$
- $0.38 < p.tox < 1$
- $0.66 < cutoff.eli < 0.89$
- $0.306 - 0.85 * target.DLT.rate < p.saf < 0.0002 + 0.9 * target.DLT.rate$
- $p.saf > 0.4295 - 2.11 * target.DLT.rate + 3.1171 * target.DLT.rate^2$
- $0.725 - 1.2 * target.DLT.rate < p.tox < 1$
- $cutoff.eli > 1.085 - 1.1141 * target.DLT.rate - 1.1438 * target.DLT.rate^2$
- $cutoff.eli < 1.0215 - 0.00858 * target.DLT.rate - 4.11 * target.DLT.rate^2$

These conditions are also visualized in **Figure 7**.

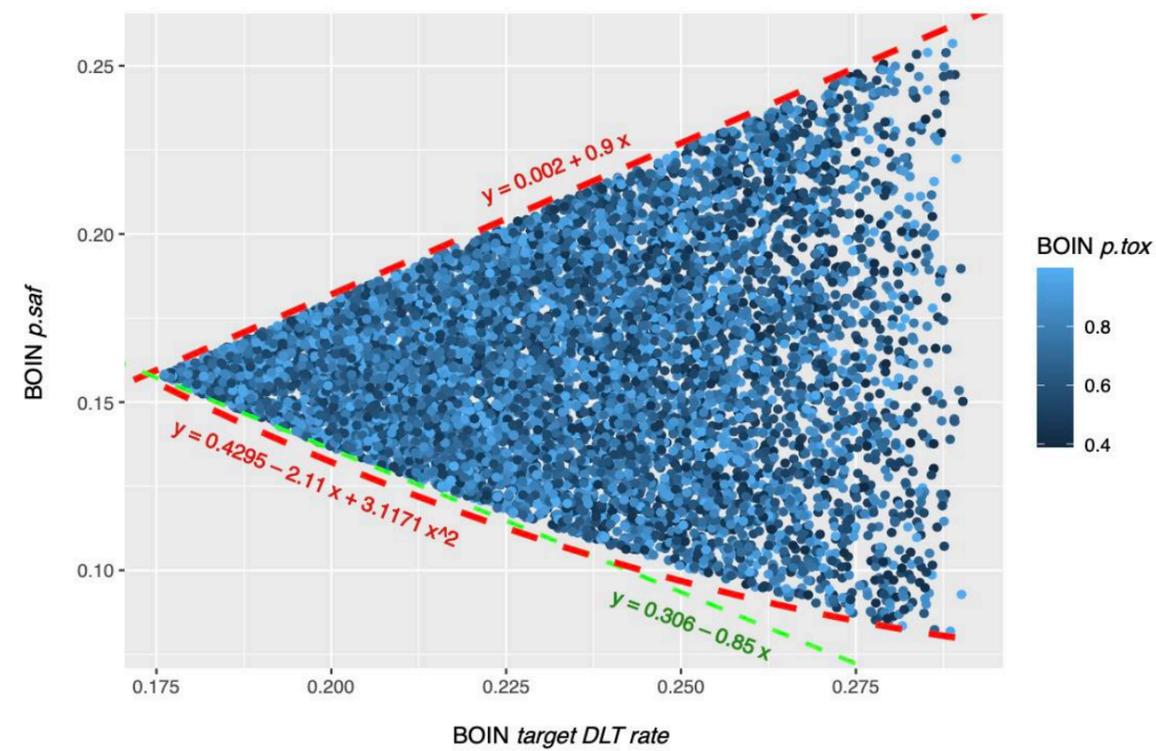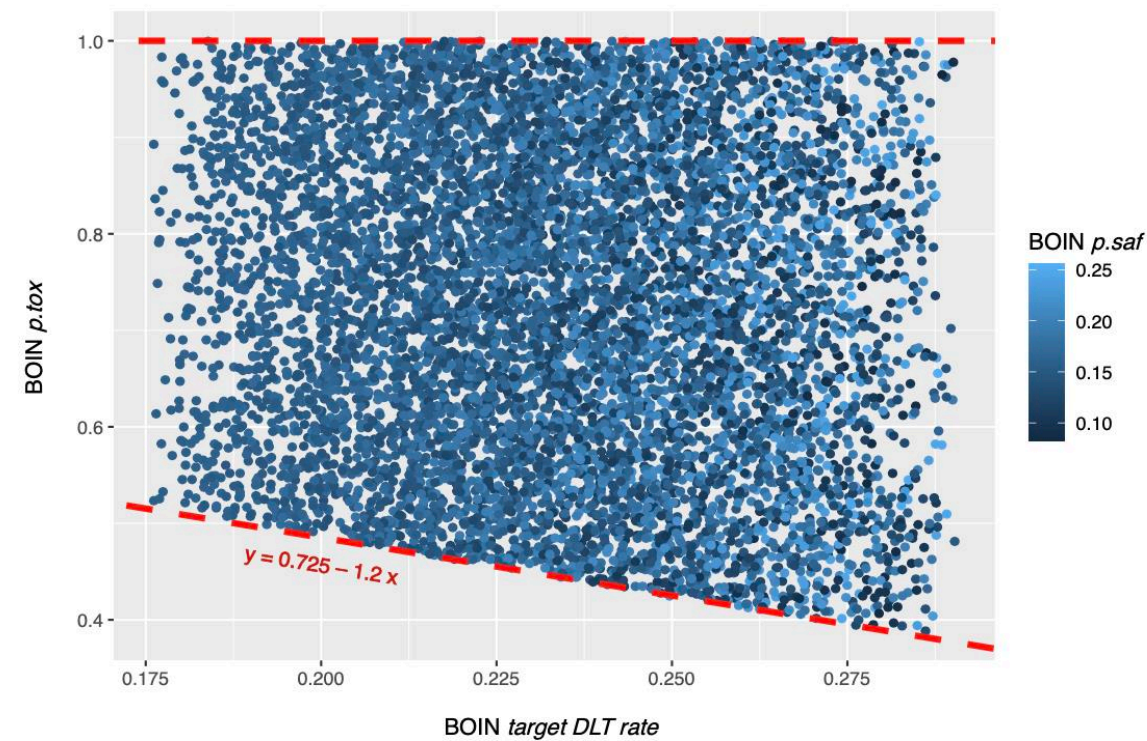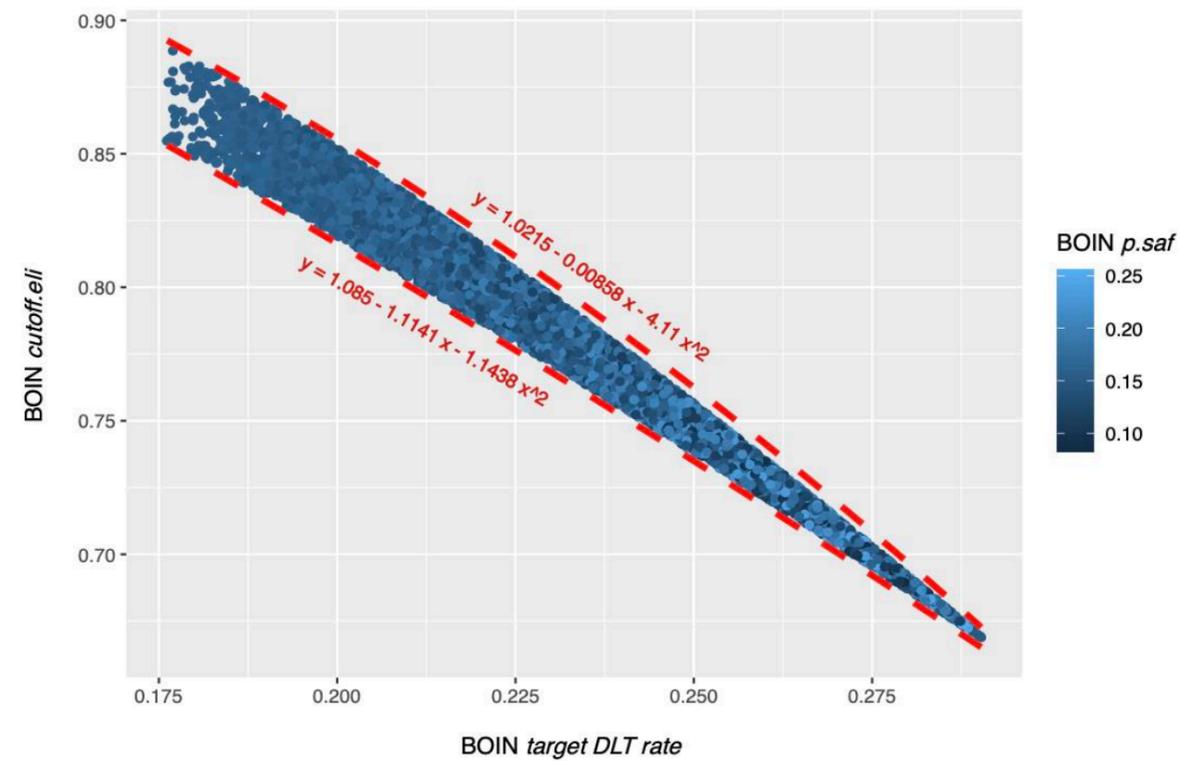

**Figure 7:** To generate the 3+3 design boundary table shown in Table 1, BOIN parameters *target.DLT.rate, p.saf, p.tox* and *cutoff.eli* need to satisfy conditions visualized above.

**DISCUSSION**

It seems that, in addition to target DLT rate, equivalent intervals of *p.tox* also depend heavily on *cohortsize, cutoff.eli* and *n.earlystop*. Although *ncohort* is one of the required input parameter of *get.boundary()*, the choice of *ncohort* value has no effect on the calculation of BOIN boundary table. The main purpose of specifying *ncohort* is to terminate dose finding process when the sample size budget is reached. When the early stopping parameter *n.earlystop* is relatively small or the *cohortsize* value is not optimized via simulation, it might be better to use *p.tox* < 1.4 * *target.DLT.rate*, or try out different cohort sizes, or increase *n.earlystop*, whichever is both feasible and provides better operating characteristics.

While changing target DLT rate, *cohortsize,* and *n.earlystop* will affect the equivalent intervals for both *p.saf* and *p.tox*, increasing or decreasing *cutoff.eli* will only affect *p.tox* equivalent intervals. It appears that increasing *cutoff.eli* will add more equivalent interval boundary points from both side of *p.tox* = 0.5 but won't be able to narrow *p.tox* intervals that are either close to target DLT rate (plus a small margin) or close to 1. And *cutoff.eli* < 90% may need to be used with caution because the resulted equivalent interval of *p.saf* could be too wide for some pediatric trials. For example, when target DLT rate = 10%, *cohortsize* = 3, *ncohort* = 10, *n.earlystop* = 12, *cutoff.eli* = 80%, and *extrasafe* = FALSE, using *p.tox* = 1.4 * *target.DLT.rate* to calculate BOIN boundary table is equivalent to using any *p.tox* ∈ (12.3%, 99.9%), as long as p.saf values fall into one equivalent interval of *p.saf*. When target DLT rate = 10%, *cohortsize* = 3, *ncohort* = 10, *n.earlystop* = 12, *cutoff.eli* = 70%, and *extrasafe* = FALSE, using *p.tox* = *1.4 ***

*target.DLT.rate* to calculate BOIN boundary table is equivalent to using any valid values of p.tox ∈ (11%, 99.9%), as long as *p.saf* values fall into one equivalent interval of *p.saf*.

Following BOIN designs may also need to be used with caution because the resulted equivalent interval of *p.saf* could be too wide for some pediatric trials:

- When target DLT rate = 20%, *cohortsize* = 3, *ncohort* = 10, *n.earlystop* = 12, *cutoff.eli* = 95%, and *extrasafe* = FALSE, using *p.tox* = 33% to calculate BOIN boundary table is equivalent to using any *p.tox* ∈ (30.5%, 48.8%), as long as *p.saf* values fall into one equivalent interval of *p.saf*.
- when target DLT rate = 25%, *cohortsize* = 3, *ncohort* = 10, *n.earlystop* = 12, *cutoff.eli* = 95%, and *extrasafe* = FALSE, using *p.tox* = 1.4 * *target.DLT.rate* to calculate BOIN boundary table is equivalent to using any *p.tox* ∈ (27.5%, 42.4%), as long as *p.saf* values fall into one equivalent interval of *p.saf*.
- When target DLT rate = 30%, *cohortsize* = 3, *ncohort* = 10, *n.earlystop* = 12, *cutoff.eli* = 95%, and *extrasafe* = FALSE, using *p.tox* = 1.4 * *target.DLT.rate* to calculate BOIN boundary table is equivalent to using any *p.tox* ∈ (36.8%, 54%), as long as *p.saf* values fall into one equivalent interval of *p.saf*.
- When target DLT rate = 10%, *cohortsize* = 4, *ncohort* = 10, *n.earlystop* = 12, *cutoff.eli* = 95%, and *extrasafe* = FALSE, using *p.tox* = *33%* to calculate BOIN boundary table is equivalent to using any *p.tox* ∈ (25.2%, 45.8%), as long as *p.saf* values fall into one equivalent interval of *p.saf*.

- When target DLT rate = 10%, *cohortsize* = 5, *ncohort* = 10, *n.earlystop* = 12, *cutoff.eli* = 95%, and *extrasafe* = FALSE, using *p.tox* = 1.4 * target.DLT.rate to calculate BOIN boundary table is equivalent to using *p.tox > 3 * target.DLT.rate*

- When target DLT rate = 10%, *cohortsize* = 6, *ncohort* = 10, *n.earlystop* = 12, *cutoff.eli* = 95%, and *extrasafe* = FALSE, using *p.tox* = 33% to calculate BOIN boundary table is equivalent to using *p.tox* = 99%, as long as *p.saf* values fall into one equivalent interval of *p.saf*.

This study demonstrates the importance of interpreting BOIN design parameter *p.tox* as an interval of toxicity rates that are considered too toxic, rather than one prespecified value that corresponds to the lowest toxicity probability that is deemed overly toxic. When designing a dose-finding trial using BOIN, it is important to perform simulation studies to identify equivalent sets of BOIN design parameters that can generate the same boundary table so that we can better compare the safety properties of different boundary tables.